\documentclass[reprint,
amsmath,amssymb,aps,
]{revtex4-2}

\usepackage{amsmath,amssymb}
\usepackage[utf8]{inputenc} 
\usepackage[T1]{fontenc}  
\usepackage{hyperref}    
\usepackage{url}      
\usepackage{booktabs}    
\usepackage{amsfonts}    
\usepackage{nicefrac}    
\usepackage{microtype}   
\usepackage{cleveref}    
\usepackage{lipsum}     
\usepackage{graphicx}
\usepackage{natbib}

\begin{document}
\title{Imaging below the camera noise floor with a homodyne microscope}

\author{O. Wolley$^{1}$, S. Mekhail$^{1}$, P.-A. Moreau$^{2,3}$, T. Gregory$^{1}$, G. Gibson$^{1}$, G. Leuchs$^{4,5}$, and M.J. Padgett$^{1*}$ \\ $^{1}$School of Physics and Astronomy, University of Glasgow, Glasgow, United Kingdom
\\ $^{2}$Department of Physics, National Cheng Kung University, No.1, University Road, Tainan City 701, Taiwan \\ $^{3}$ Center for Quantum Frontiers of Research and Technology, NCKU, Tainan 70101, Taiwan 
\\ $^{4}$Max Planck Institute for the Science of Light, Staudtstr. 2, D-91058 Erlangen, Germany
\\ $^{5}$Department of Physics, Friedrich-Alexander-Universität Erlangen-Nürnberg, Erlangen, Germany\\ $^{*}$miles.padgett@glasgow.ac.uk}


\date{\today}
\begin{abstract} 
Low-light imaging can be challenging in regimes where low noise detectors are not yet available due to image contrast being degraded by sensor noise. We present a homodyne imaging system capable of recovering intensity and phase images of an object from a single camera frame at an illumination intensity significantly below the noise floor of the camera. By interfering a weak imaging signal with a reference beam that is $\sim\!300$,$000$ times brighter, we are able to image objects in the short-wave infrared down to signal intensity of $\sim\!1.1$ photons per pixel per frame incident on the sensor despite the camera having a noise floor of $\sim\!200$ photons per pixel. There is a corresponding $29.2\%$ drop in resolution of the image due to the method implemented. We believe our demonstration could vastly extend the range of application for low-light imaging by accessing domains where low noise cameras are not available.
\end{abstract}
\maketitle
\section{Introduction} 
Techniques to perform imaging under a low-photon flux are important in various contexts, such as medical and biological imaging where radiation exposure has to be minimised and covert imaging where a low-photon flux is required to avoid detection. As the illumination level is reduced the electronic readout noise of the detector becomes an important consideration and noticeably degrades resulting image quality. Once the detected signal is of the order of the detector noise it is no longer possible to distinguish the object being imaged from the background. Whilst detector noise is typically equivalent to a single photon per pixel per frame for modern silicon based array detectors designed to operate at visible wavelengths, detector noise is much more significant at infrared wavelengths, such as short-wave infrared (SWIR) regions. Here, non-silicon based detectors such as gallium arsenide (GaAs) are used which have noise floors at least an order of magnitude higher. Increasing the illumination level mitigates high detector readout noise, however, this is not always possible in low-light imaging contexts in which photon flux must be minimised. For example, in the context of biological imaging applications, where illuminating a live sample with a high probe beam power may alter cell processes and high probe beam power can lead to cell death \cite{taylorQuantumMetrologyIts2016,khodjakovImagingDivisionProcess2006}. In absorption measurements increasing the probe power can also lead to saturation of the sample \cite{bieleMaximizingPrecisionSaturationlimited2021}. Further to these applications, low-light imaging contexts exist within LiDAR and covert imaging where, either low-power eye-safe sources are required, or in cases where imaging is performed through a scattering media \cite{Tobin:19, Maccarone:15}. 

Various interference based methods have been demonstrated as ways of performing sensitive measurements. Non-linear techniques such as frequency upconversion and non-linear interferometry are capable of recording sensitive measurements at infrared wavelengths on visible detectors \cite{Boyd, Zeilinger}. Homodyne and heterodyne detection have also been widely used to measure with a sensitivity below the noise constraints of a system across various applications \cite{JinHeterodyne, Caves, MasonForce, PhysRevA.86.032106, Lobanov:14, SCHIEDER1994477, 1974A&A....36..341V}. Using homodyne measurements, where the interference of two waves originating from the same source is measured rather than a direct signal, one can measure with a greater sensitivity and have access to both amplitude and phase information of an unknown wavefront. This can be done since the interference allows for the recovery of phase dislocations in the signal wavefront\cite{Nye:74}. As such homodyne measurements also allow for measuring phase singularities \cite{Dorn:2000}, potentially leading to super resolution imaging\cite{TYCHINSKY1991131}. In this work, we concentrate on the optical amplifying property of homodyne measurements to surpass the noise floor of an imaging detector.

One form of optical homodyne measurement is digital holographic microscopy. A technique in which a hologram \cite{gaborNewMicroscopicPrinciple1948} is created by interfering the probe beam, which interrogates the object of interest, with a reference beam. The resulting interference pattern is imaged onto a digital sensor \cite{goodmanDIGITALIMAGEFORMATION1967,kimPrinciplesTechniquesDigital2010}. By recording the interference pattern, both the intensity and phase of the object can be retrieved from a single frame. In a balanced interferometer the intensity of the probe and reference beams are equal and the contrast of the resulting interference fringes is maximised. However, if the probe and reference beams are unbalanced, an interesting effect occurs due to the relationship between electric field and intensity, as articulated by Gabor \cite{GABOR1961109}. If the reference to probe amplitude ratio is set to $100$:$1$, the resulting interference fringes still have a contrast of $\approx\!10\%$ due to the fact that the change in amplitude of the field results in a squared change of the recorded intensity. This means that small changes in the intensity of the probe beam will lead to large changes in the resulting interference. In this sense, the homodyne measurement can be treated as an optical amplification process. Considering $I_{\text{tot}}$, the intensity profile of the interference image, 
\begin{equation}
  I_{\text{tot}} = I_{\text{ref}} + I_{\text{probe}} + 2(I_{\text{ref}}I_{\text{probe}})^{\frac{1}{2}}\cos{\phi},
\end{equation} 
where $I_{\text{ref}}$ and $I_{\text{probe}}$ are the intensities of the reference and probe respectively and $\phi$ the relative phase between the two arms. The term $(I_{\text{ref}}I_{\text{probe}})^{\frac{1}{2}}$ gives the enhancement of a weak probe signal. This benefit of interference allows imaging below the noise floor of the detector in a classical imaging system.

A homodyne detection scheme has been implemented in the context of quantum imaging in which squeezed states of light were used to probe an object and image below the noise floor of the detector by interfering a signal with a local oscillator \cite{cuozzo_new}. The authors have also generalised their system to work with thermal states, obtaining shadow images of a semi-transparent object by measuring the mean temporal variance between the two output ports of the interferometer \cite{Barge_2022}. Using a reference hologram obtained with the probe beam at a high light level to imitate the interference pattern when imaging at a low-light level has been performed to enable imaging for "largely invisible and noisy" interference fringes \cite{inoueAngularSpectrumMatching2021}. That method uses angular spectrum matching and requires two exposures to obtain the reference (object out, high-light) and image (object in, low-light) holograms. The method of using an unbalanced interferometer has also been used to measure orbital angular momentum modes of light in a low illumination regime \cite{ariyawansaAmplitudePhaseSorting2021b}. Whilst we are unable to image at as low photon counts as a quantum scheme, our classical imaging technique is capable of real-time imaging from a single frame with the probe beam below the noise floor of the detector without the requirement of a quantum source. Furthermore, we obtain both intensity and phase images of an object allowing for greater sample inspection. This unbalanced homodyne imaging gives us the ability to image the object of interest using a low power probe beam whilst retaining the ability to recover an image not dominated by detector noise by using a high power reference beam. Such a scheme has the potential to be operational all the way down to the single photon regime, as has previously demonstrated for the case of two pixels \cite{hessmo_experimental_2004}. 

This method of performing a homodyne measurement to extract both the intensity and phase information of a sample has been applied in biological imaging for cell imaging and disease identification \cite{Kemper:08,Anand2017AutomatedDI}. Label-free biological imaging methods have the advantage of not being chemotoxic or causing modifications to cell chemistry which can kill the sample or interfere with observations of cell processes \cite{waldchenLightinducedCellDamage2015}. The method could be further adapted to non-destructive testing and inspection applications that require imaging at low illumination levels or at wavelengths for which detectors have a high noise floor due to readout and environmental noise such as is the case for detectors operating in the far-infrared or terahertz. Furthermore, in the context of medical imaging, patient dose is a important consideration so applying this technique to medical imaging schemes to amplify a weak signal that has been used to probe the patient with a strong local oscillator in order to limit patient dose. Other diffractive imaging techniques, such as Fourier ptychography, could also serve in some of these contexts for the purpose of obtaining phase images \cite{aidukasPhaseAmplitudeImaging2019,wojdylaEUVPhotolithographyMask2018}. Further interference based imaging schemes, such as those that utilise multiple wavelengths, could also benefit from this technique \cite{tobinOpticalSpatialHeterodyned2006,khmaladzePhaseImagingCells2008}.

In this work we present a homodyne detection system that is capable of imaging under detected probe beam illumination levels that are below sensor readout noise in the short wave infrared wavelength regime. We are able to recover intensity and phase images where features of the object are distinguishable down to a signal intensity of $\sim\!1.1$ photons per pixel incident upon the camera sensor. This corresponds to a reference beam to probe beam power ratio of $\sim\!300$,$000$:$1$. We are able to recover a measurable image contrast for the homodyne interference image as compared to a wide-field image where noise dominates over the image contrast. This enhancement in image contrast for the homodyne image relative to the direct image has a corresponding reduction in image resolution of $29.2\%$. Furthermore, phase images are also recovered which allows depth measurements to be made. We believe our demonstration shows the possibility of low-light imaging in domains where detector noise is a significant issue, extending the range of low-light imaging applications. 

\section{Methods}
\subsection*{Imaging system}
\begin{figure*}[t]
	\centering
	\includegraphics[width=0.9\linewidth]{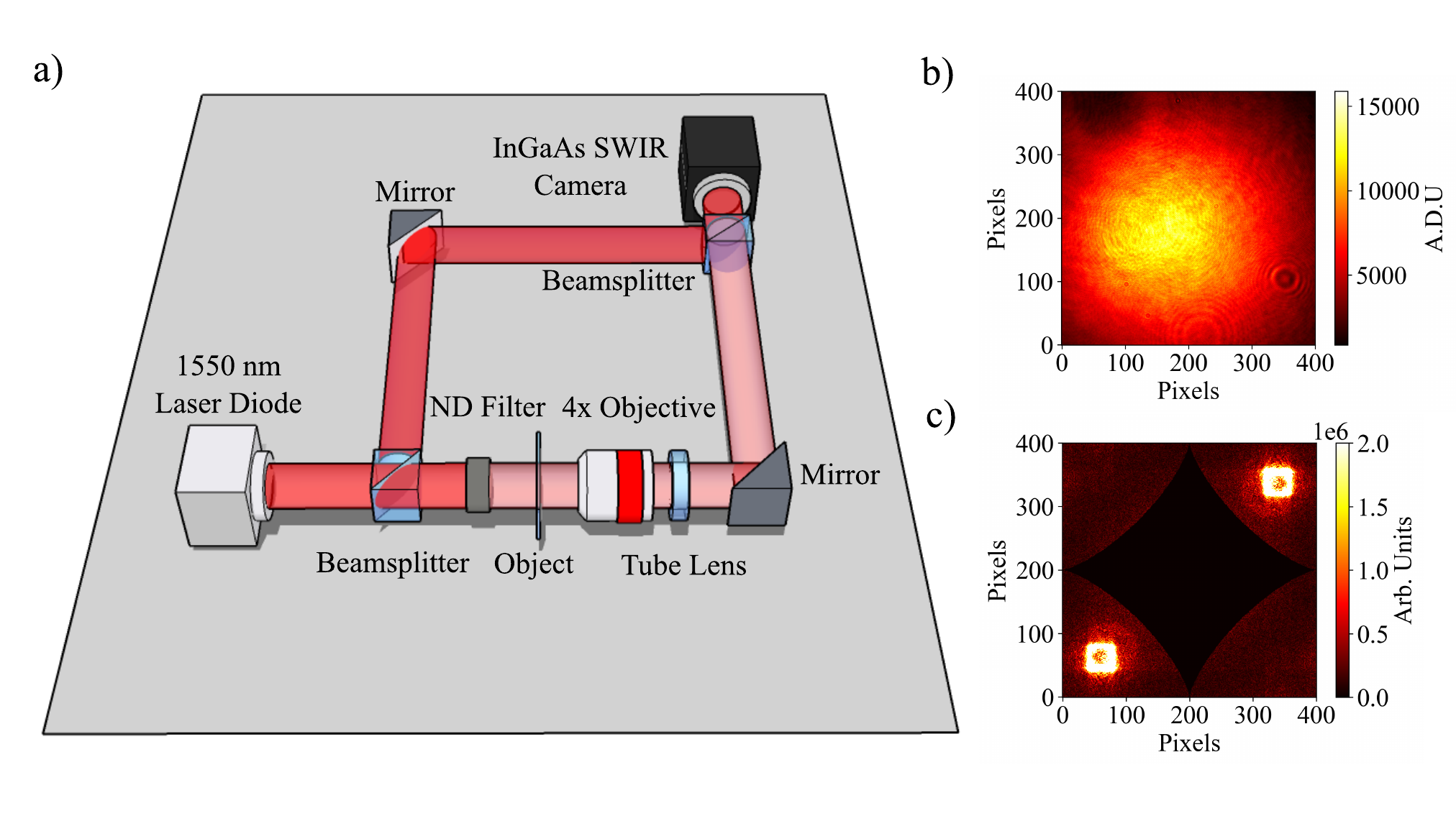}
	\caption{ a) Diagram of the experimental method setup to obtain interference images. We use a microscope built in a transmission configuration using a $4\text{x}$ microscope objective and $150\text{ mm}$ tube lens to image the object onto the camera. To perform an optical homodyne measurement we split our illumination source using a beamsplitter. The signal beam passing through the object is attenuated to a low intensity by the use of a neutral density (ND) filter, before being interfered at the second beamsplitter with a reference beam of much greater intensity. b) Example hologram recorded on the detector, showing low fringe contrast due to the unbalancing effect. c) Example of a FFT of an obtained hologram, with the central DC components masked out by a curved diamond shape.}
	\label{fig:setup}
\end{figure*}
An inverted wide-field microscope was built with a secondary reference arm as presented in the experimental setup in Figure \ref{fig:setup}. The laser used is a $1550 \text{~nm}$ laser diode with a coherence length of $24.0\text{~mm}$. To perform an optical homodyne measurement the laser beam is split into the signal arm and the reference arm at the first beamsplitter. The object is placed into the signal arm and the relative intensity of the two beams is controlled using reflective neutral density (ND) filters, with a more optically dense filter in the signal beam such that the reference beam has a higher intensity. The signal and reference beams are then recombined at a second beamsplitter and the object imaged onto the camera detector array. 

The camera used is a SWIR camera, with a typical readout noise of $180e^{-}$ per pixel and quantum efficiency (QE) of $85\%$ at $1550 \text{ nm}$. Taking this QE into account, the effective noise floor of the camera is $\sim\!207$ photons per pixel. We use the low gain mode of the sensor, which while noisier, gives a higher dynamic range to allow for a greater ratio between signal and reference beams.

\subsection*{Intensity and Phase Reconstruction} 
We obtain an intensity image of the interfered signal and reference beams on our camera, which contains information about the signal and reference fields $E_{\text{sig}}$ and $E_{\text{ref}}$ according to 
\begin{equation}
  \begin{aligned} 
I_{\text{tot}} &\propto |E_{\text{sig}}(\mathbf{r}) + E_{\text{ref}}(\mathbf{r})\text{exp}(i \mathbf{k_{tilt} r}) |^2\\
 &= I_{\text{sig}} + I_{\text{ref}} + 2\text{Re}[ E_{\text{sig}}(\mathbf{r})E_{\text{ref}}(\mathbf{r})\text{exp}(i \mathbf{k_{tilt} r})],
\end{aligned}
\end{equation} 
where $\mathbf{k_{tilt}}$ is the relative wavevector between the propagation of $E_{\text{sig}}$ and $E_{\text{ref}}$. We take an average of the reference beam over a minimum of a hundred frames, and subtract this from the hologram. $E_{ref}$ is assumed to have a flat phase front which is fixed by proper alignment of the reference beam and the field is therefore estimated as $\sqrt{I_{ref}}$. In a method similar to that of Fatimi and Beadie \cite{fatemiRapidComplexMode2013}, we take a FFT of $I_{tot}-I_{ref}$. Since $I_{sig}\ll I_{ref}$ and is also below the noise floor of the camera it is simply neglected. Any remaining low spatial-frequency components of the FFT are masked as per Fig.~\ref{fig:setup}~(C). One of the four quadrants is selected by the user to highlight the location of the interference pattern and the other three are masked. For example, as can been seen in Fig.~\ref{fig:setup}~(C), two bright areas are present in diagonally opposing quadrants. The selection of the wrong one simply gives the conjugate phase of the object being imaged. Determining which quadrant corresponds to the relative tilt between the reference and probe arms can be done with a reference sample. Once the unused quadrants are masked the relative wavevector $\mathbf{k_{tilt}}$ is determined by an algorithm based on a centre of mass calculation. This is used to approximate the relative reference field as $\sqrt{I_{ref}}e^{i\mathbf{k_{tilt}r}}$. The masked FFT then undergoes an inverse FFT and the resulting complex field is point-wise coherently divided by the relative reference field. The result is the complex field image of the sample which is then split into modulus and argument to give the transmissive amplitude and phase profile of the sample, respectively. 

We confirmed the linearity of the greyscale between a conventional image and a homodyne intensity image by imaging a ND filter where half the frame was covered by the filter and the other half clear glass and ensuring the greyscale values between the two regions dropped by an equivalent amount for both the conventional wide-field image and homodyne image.

\begin{figure*}[ht]
	\centering
	\includegraphics[width=0.95\linewidth]{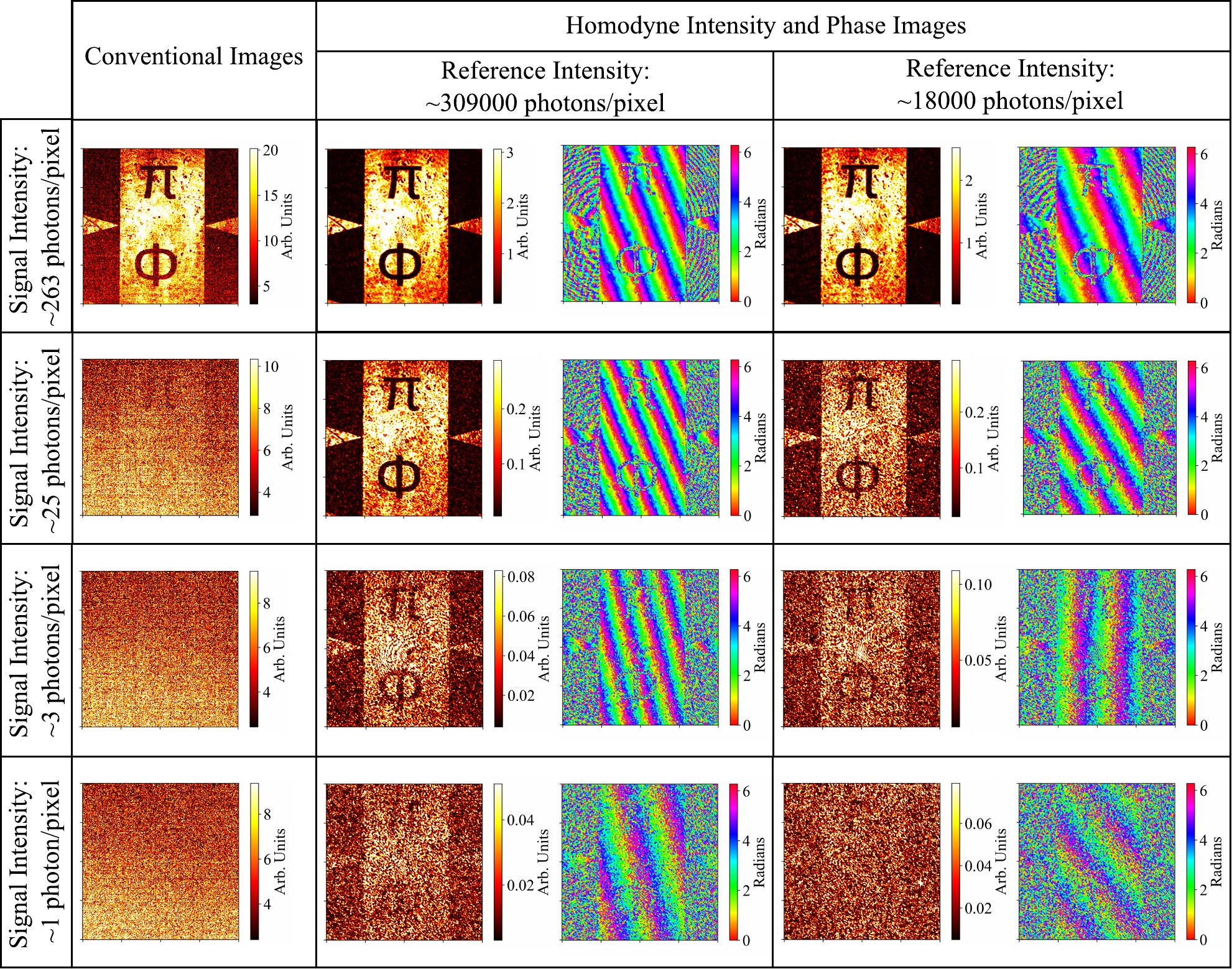}
	\caption{A table comparing conventional camera images and reconstructed homodyne intensity and phase images at differing illumination levels. Shown are conventional camera images at three different signal intensities incident on the camera alongside the corresponding reconstructed intensity and phase images from the homodyne system of a silicon chip with gold deposited features. The reconstructed images are shown using two different reference beam intensities. Due to spatial filtering inherent in the homodyne image reconstruction process, the conventional images have had a background noise subtraction and equal degree of spatial filtering applied for fair comparison. All conventional and homodyne intensity images were normalised by setting the minimum and maximum of the scale to the value of the $10^{\text{th}}$ and $90^{\text{th}}$ percentile of pixel values respectively. The presence of fringes in the phase images arise in the phase correction process due to the varying thickness of the silicon chip.}
	\label{fig:homodynetable}
\end{figure*}

Due to physical constraints of the microscope the probe arm introduces a spherical phase front not present in the reference arm. As such this is measured and subtracted from the calculated phase profile of the sample. Due to the instability of the system on a frame by frame basis, we apply temporal averaging on our reference phase image such that errors are not incurred during the subtraction. This averaging occurs during the measurement of $I_{ref}$ the sample is placed in the system so no excess illumination of the sample is required.

\section{Results}
\begin{figure*}[ht]
	\centering
	\includegraphics[width=0.65\linewidth]{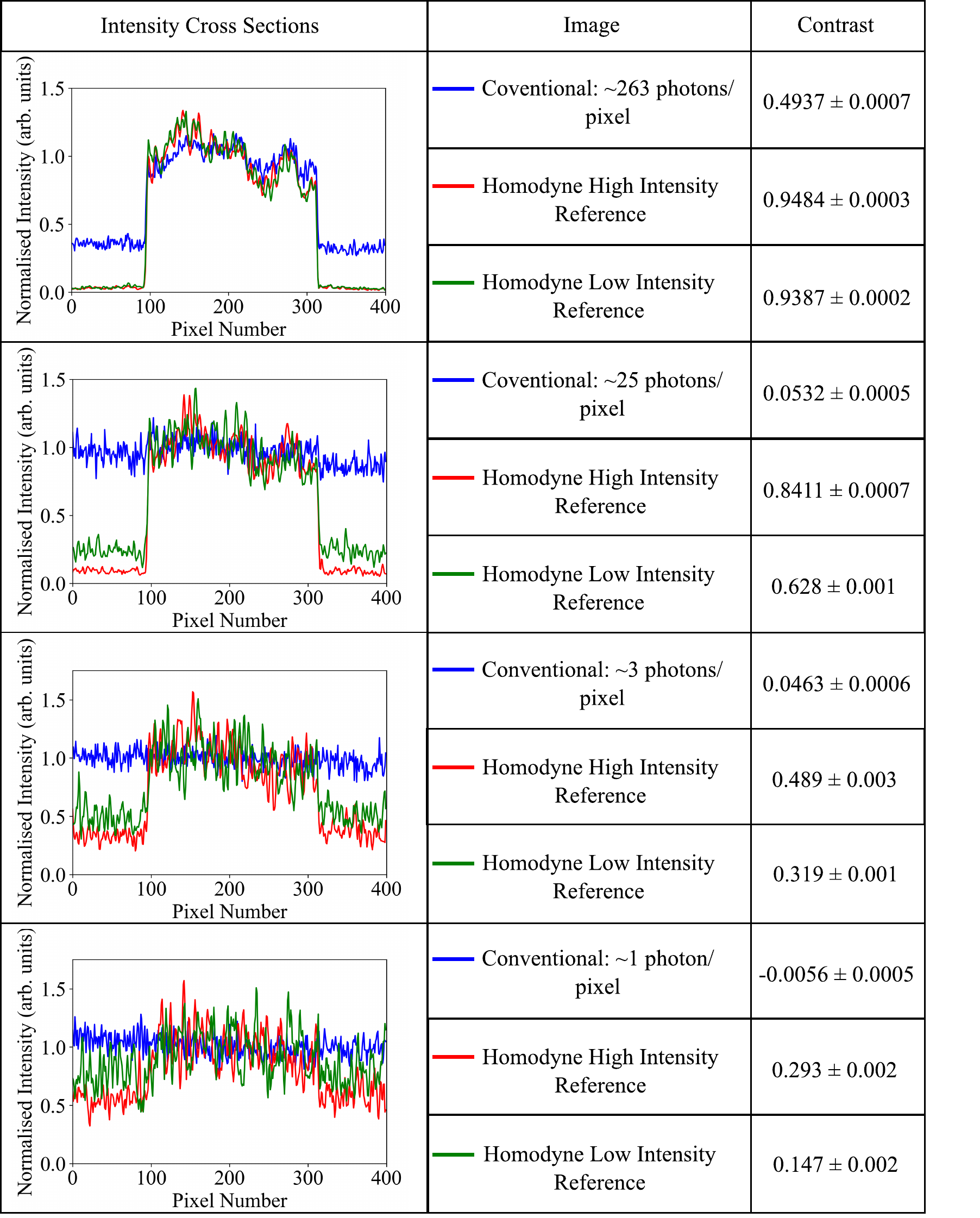}
	\caption{Image intensity cross-sections for the images displayed in Fig. \ref{fig:homodynetable} alongside image contrast statistics. Image intensity cross-sections were produced by averaging image intensity over columns between two sets of rows for each image (conventional camera image, homodyne intensity at a high intensity reference and homodyne intensity at a lower intensity reference). Columns $0-400$ were averaged over between rows $10-45$ and $230-265$ to generate the intensity cross-section line plots. These plots were displayed on the same scale by normalising each to the average value of the bright regions of the cross-section, columns $100-300$. The contrast was then determined using the average value of these bright regions and the average value of the dark regions and averaged over 100 frames with the standard error on the mean calculated. The dark regions were determined by columns $0-75$ and $325-400$. }
	\label{fig:homodynecuttrhoughs}
\end{figure*}

A series of images of a silicon chip with gold deposited features acquired by the system under a range of different illumination scenarios is shown in Fig. \ref{fig:homodynetable}. The silicon regions are transparent to light in the SWIR whilst the gold regions reflect the light. Conventional images of the signal beam are shown at three different illumination levels, along with homodyne intensity and phase images reconstructed from holograms recorded at two different reference beam intensities. As the reconstruction of intensity and phase images involves spatial filtering (masking in the Fourier plane) of the DC components, we perform a background noise subtraction and apply an equivalent amount of spatial filtering for fair comparison. All images shown were recorded in real time from a single camera frame. We calculate the intensity of the reference beam detected on the camera by converting the pixel values recorded into a number of photons using the stated full well capacity of the pixels and quantum efficiency of the sensor at $1550\text{~nm}$. We are unable to use this method to calculate the signal intensity at the camera, as at lower illumination levels the signal information is contained within one bit, so we cannot convert accurately to a photon number. Instead, we calculate the ratio between the two arms and use the calculated reference intensity to obtain a value for the intensity of the signal arm. In order to do this, we use the manufacturer stated transmission at $1550 \text{~nm}$ of the ND filters used, and measure the drop in power across the objective and tube lens using a power meter.

It can be seen from Fig. \ref{fig:homodynetable} that the homodyne system is capable of recovering contrast of the object in the signal arm, either by reducing noise on the image when the object is partially visible or even recovering features of the object when the image contrast is dominated by camera noise. In the bottom row of Fig. \ref{fig:homodynetable} we present an image obtained for a single frame with an estimated probe intensity of $1.1$ photons per pixel per frame. This means for an effective noise floor of $\sim\!207$ photons per pixel we are able to obtain intensity and phase images from a single camera frame with no requirement to perform pixel binning in post processing under an illumination level $\sim\!200$ times below the detector noise.
\begin{figure*}[ht]
	\centering
	\includegraphics[width=\linewidth]{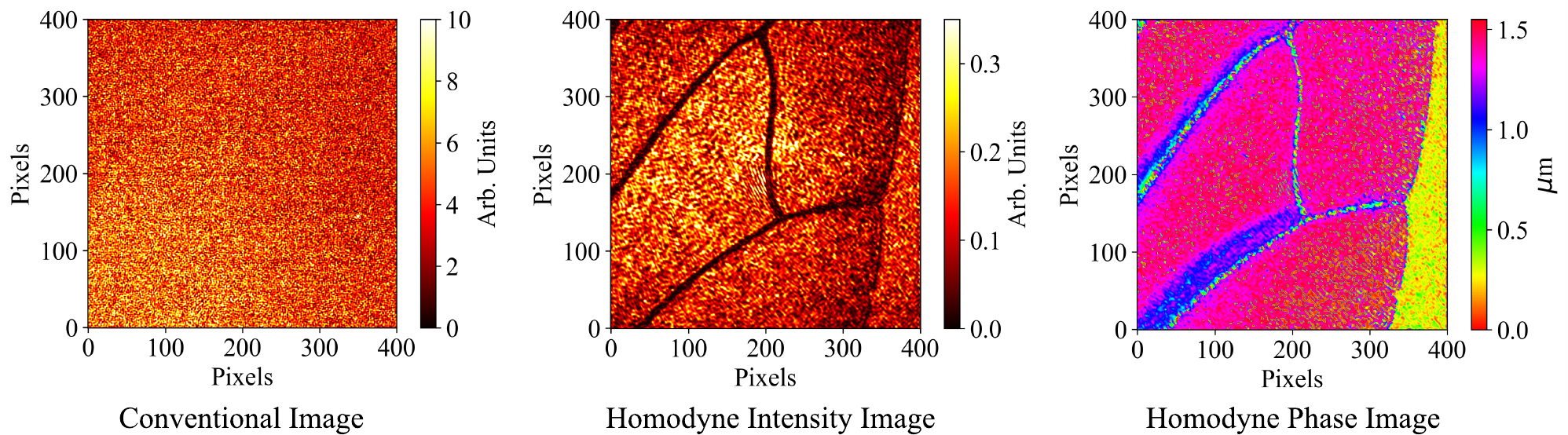}
	\caption{Single frame conventional camera image alongside corresponding homodyne intensity and phase images of an insect wing. Homodyne intensity and phase images show features of a transmissive object can be recovered by the system when they are not visible in a conventional camera image. Furthermore, the access to phase information allows for depth information of the sample to be obtained by a conversion of the scale. It can be seen that the wing has an associated thickness which is greater towards the base of the wing. The intensity of the signal incident on the camera was calculated to be $\sim\!25$ photons per pixel.}
	\label{fig:wingphase}
\end{figure*}

The performance of the system is linked to the intensity of the reference beam. If the intensity of the reference beam is lowered, the signal is not amplified to as greater an extent and the recovered homodyne images are noisier and have reduced contrast. This can be seen in Fig. \ref{fig:homodynetable} and is further illustrated in Fig. \ref{fig:homodynecuttrhoughs} showing intensity cross-section plots of the images in Fig. \ref{fig:homodynetable}, alongside calculated values of the contrast for each image. Intensity cross-sections for the homodyne and signal images are taken by averaging over columns between two sets of rows with clear bright and dark regions, rows $10-45$ and $230-265$. They are then displayed on the same plot by normalising to the average value of the bright regions in each cross-section. The contrast was calculated as $(I_{\text{bright}} - I_{\text{dark}})/(I_{\text{bright}} + I_{\text{dark}})$, where $I_{\text{bright}}$ and $I_{\text{dark}}$ are the average intensity values for the bright and dark regions of the image respectively. This was averaged over 100 frames with the standard error on the mean also calculated. It can be seen particularly at the lower signal intensities that image contrast is reduced for a lower intensity reference beam, due to the interference term in the intensity profile recorded on the camera not being amplified to as greater extent. At these lower reference beam intensities the detector noise becomes significant and the images obtained have reduced contrast, demonstrating the importance of a high powered reference to preserve image contrast. 

It can also be seen from Fig. \ref{fig:homodynecuttrhoughs} that as the signal intensity decreases, the contrast of the recovered homodyne images decreases. Whilst the homodyne method is able to decrease the contributions of detector noise, it cannot eliminate shot noise arising from the fluctuations of light itself. These fluctuations will make a contribution in the subtraction of the reference beam as well as on the signal itself. As the intensity of the signal decreases, the signal to noise ratio decreases and there is reduced contrast in the recovered homodyne images. At an illumination level of $\sim\!1.1$ photons per pixel whilst the contrast is reduced to $0.293\pm0.002$, the contrast in the corresponding conventional image is completely dominated by noise, measured as being $-0.0056\pm0.0005$. This means features of the object can be distinguished using the homodyne imaging method which would otherwise not be observable with simple conventional imaging.
\begin{figure*}[ht]
	\centering
	\includegraphics[width=0.7\linewidth]{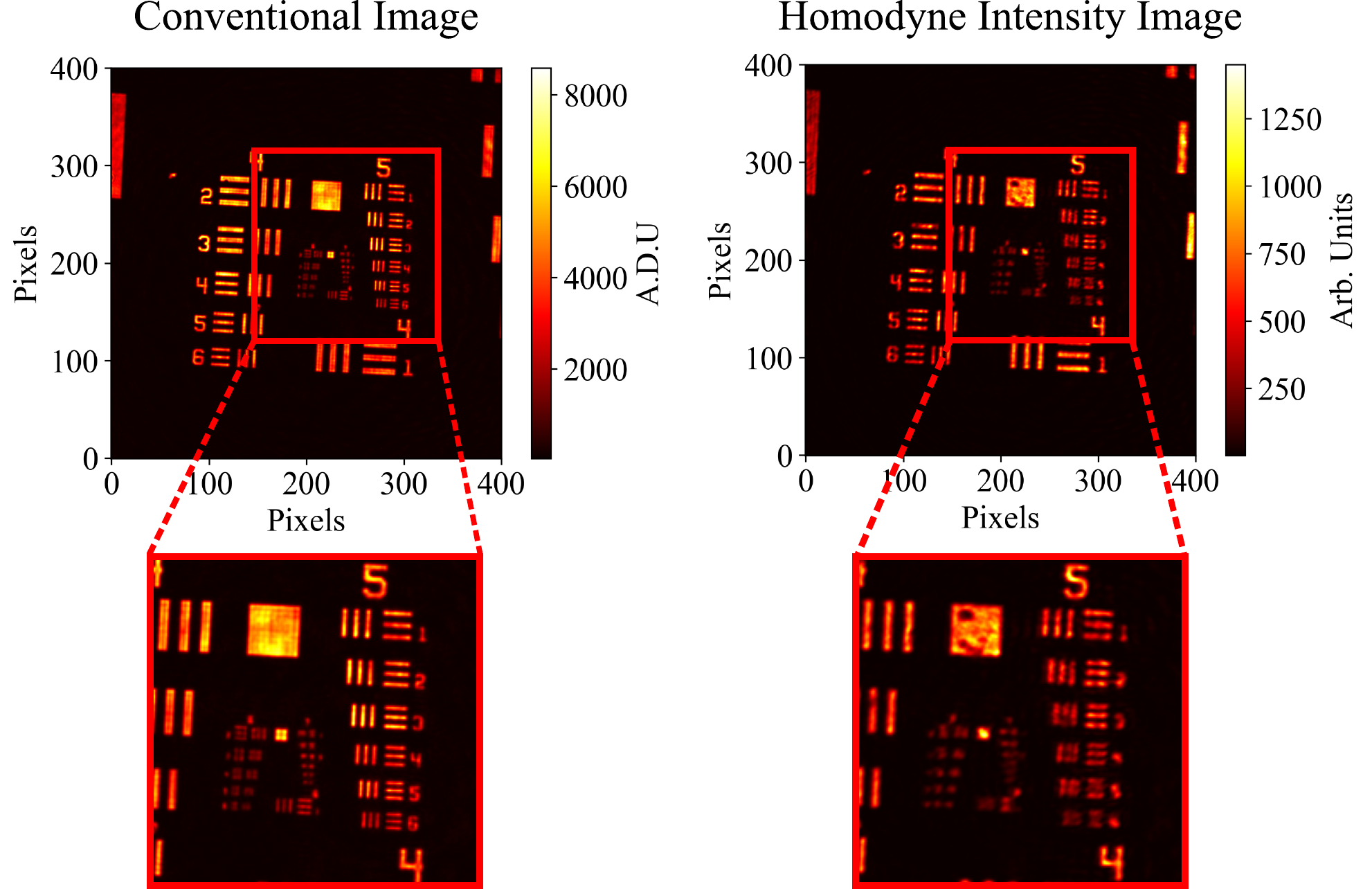}
	\caption{Images of a USAF resolution target from the conventional camera image and reconstructed homodyne intensity image.It can be seen that in the conventional camera image the smallest resolvable element is $1$ in group $6$ (largest element in the lower left of inner square), whilst for the homodyne image it is element $4$ of group $5$ (4th element down on right hand side). This corresponds to a loss in resolution of $29.2\%$ when images are reconstructed using the homodyne method when compared with what could be achieved with a conventional camera image.}
	\label{fig:resolution}
\end{figure*} 

We also present the imaging of a transmissive object at a low illumination level, and the recovery of phase and depth information from the sample. Fig. \ref{fig:wingphase} shows an image of an insect wing at an illumination level of $\sim\!25$ photons per pixel, from a single camera frame. Shown is the direct signal image, along with intensity, phase and depth information recovered from the hologram. From the phase image it is possible to convert the scale into depth information in $\mu \text{m}$. Access to depth information allows for greater inspection of a transmissive object, as it can be seen from the scale in the phase image that the wing gets thinner towards the vein areas. The wrapping of the phase seen in areas of the wing is due to the thickness of the sample being greater than the wavelength used to probe it. Using multiple frequencies and using the beat frequency as shown in \cite{tobinOpticalSpatialHeterodyned2006} would reduce this wrapping effect. The ability to record phase and depth information would be attractive in the imaging of biological and material samples as it can reveal features that are otherwise not visible from an intensity profile alone at a sensitivity level not accessible without homodyne amplification. 

With our system, some resolution loss is unavoidable due to spatial filtering applied in frequency space. However, the resolution loss presented here is not an inherent limitation of the homodyne method, but of our optical system. We argue that if the wavevector $\mathbf{k_{tilt}}$ is large enough then we could sufficiently separate the information contained by the interference fringes from the central DC components in frequency space, and thus masking DC components would not lead to any spatial frequencies corresponding to image information being lost. In practise, this would be achieved by $\mathbf{k_{tilt}}$ being larger than the numerical aperture (NA) of the system at the detector, given by the NA of the microscope objective divided by the magnification. In our system, using a low magnification objective, we cannot achieve a value of $\mathbf{k_{tilt}}$ larger than the NA at the detector due to the pixel pitch of the camera. However, with a higher magnification objective or smaller pixel pitch it should be possible to obtain fringes with $\mathbf{k_{tilt}}$ larger than the NA at the detector and not observe a drop in resolution in the reconstructed homodyne images when compared with the conventional image.
In order to quantify the resolution loss of our experimental setup, a raw image of a USAF resolution target is compared with its corresponding homodyne intensity image in Fig. \ref{fig:resolution} to calculate the loss of resolution in the obtained homodyne images. It can be seen from Fig. \ref{fig:resolution} that in a conventional image taken with the system, element $1$ of group $6$ is resolvable, giving a resolution of $64.0 \text{ lp mm$^{-1}$}$ (line pairs per millimetre). We see in the corresponding homodyne image that element $4$ of group $5$ is the smallest resolvable element, giving a resolution of $45.3 \text{ lp mm$^{-1}$}$. This means with our implementation of the homodyne method there is a $29.2\%$ drop in resolution in the corresponding intensity and phase images. With some trade off between noise sensitivity and resolution, we chose a mask size which preserves noise sensitivity without losing too much resolution and used a shape of form $(x^p + y^p)^{(1/p)} = r$, which for a value of $p = 0.75$ gives a curved diamond shape. The radius of the mask used was 200 pixels. Whilst there is a modest loss in resolution, we anticipate the homodyne method being most useful in situations where features of the object would otherwise be obscured by noise and not resolvable due to required low illumination levels.

\section{Discussion}
We have presented a homodyne detection method capable of reconstructing intensity and phase images below the camera noise floor at a range of different illumination levels, with image contrast increased down to a detected signal intensity of $1.1$ photons per pixel. This shows the ability of the system to image under the conditions of low sample illumination below the noise floor of the detector. With a camera with an effective noise floor of $\sim\!207$ photons per pixel, we are able to image under the conditions of a signal intensity $\sim\!200$ times lower than the noise floor of the camera. With our imaging system a degradation in resolution of $29.2\%$ is observed, however, we note that this loss in resolution is not inherent to the method but due to the imaging system used. The system can also image transmissive objects, and recover intensity, phase and depth information about the object, being able to determine the thickness of a sample. The enhancement in sensitivity is made possible because homodyne detection amplifies the optical signal before the photoelectric conversion process at the detector.                             

We have identified the intensity of the reference beam to be a key parameter to the performance of the system, with a lower reference intensity leading to reduced contrast in recovered intensity images. As such, the dynamic range of the camera is an important consideration when designing the system. We report recovering homodyne intensity and phase images at a maximum ratio between signal and reference beams of $\sim\!300$,$000$:$1$, using a camera with a dynamic range of $72 \text{~dB}$.

With the ability to perform a wide-field imaging protocol in real time with signal intensities below the noise floor of the detector we believe our system could find applications in low-light imaging scenarios where low noise detectors are not currently available. With the system being applicable to a broad wavelength range, it could vastly expand applications in low-light imaging contexts. Here it could be used to either decrease acquisition times to obtain images of equivalent contrast at faster speeds or to allow imaging at illumination levels where it would otherwise not be possible to determine the features of an object.  

\section{Materials and Methods}The infrared light source at $1550 \text{ nm}$ is a Thorlabs LP1550-PAD2 with an actual centre wavelength of $1548.7\text{ nm}$ and a typical linewidth of $0.1\text{~nm}$, giving a coherence length of $24.0\text{ mm}$. The SWIR infrared camera is a Raptor Owl $640\text{M}$ SWIR camera with an InGaAs sensor of $15\mu\text{m} \times 15\mu\text{m} $ pixel size. The noise floor of the camera was stated to be $180e^{-}$ per pixel. With the sensor having a quantum efficiency (QE) of $85\%$ at $1550 \text{ nm}$, $1e^{-} = 1.18$ photons, we calculated the effective noise floor to be $\sim\!207$ photons per pixel. The power meter head used to determine the absolute and relative powers of the probe and reference beams in this experiment is a Thorlabs S122C. The estimated proportion of light at $1550 \text{~nm}$ that passes through the microscope objective and tube was determined to be $\sim\!56\%$ using the $1550 \text{~nm}$ laser and measuring the power before and after the optics. The sample illumination was determined by measuring the power after the first beamsplitter and the manufacturer stated transmissions of the ND filters, and using a measured beam diameter of $3 \text{~mm}$. The ND filters used were Thorlabs NEIRxxA ND filters an the manufacturer stated transmissions were used in calculating the power of the signal beam. ND filters and transmissions for each dataset in presented in Fig. \ref{fig:homodynetable} are as follows: NENIR05A-C - $34.24\%$, NENIR20A-C - $1.2818\%$, NENIR30A-C - $0.1472\%$ and NENIR40A-C - $0.0144\%$.

\section{Acknowledgements}The authors would like to thank Raptor Photonics for providing a loan of the Raptor Owl $640\text{M}$. 

\subsection*{Funding}This work was funded by the UK EPSRC (QuantIC EP/M01326X/1). O.W. acknowledges the financial support from the EPSRC (EP/R513222/1). S.M. acknowledges the EPSRC (QuantIC EP/M01326X/1). P.-A.M. acknowledges the support from the Ministry of Science and Technology, Taiwan (Grants No. 110WFA0912122). T.G. acknowledges the financial support from the EPSRC (EP/SO19472/1). G.G. acknowledges the financial support from the EPSRC (EP/T00097X/1). G.L. acknowledges financial support by BMBFunder contract No. 13N15329 ("HoChSEE"). M.J.P. acknowledges the financial support from the Royal Society (RSRP/R1/211013P).

\subsection*{Author Contributions}M.J.P. and G.L. conceptualised the project. M.J.P., T.G. and S.M. supervised the project. O.W., P.-A.M., S.M., T.G., G.G. and M.J.P. devised and implemented the project methodology. O.W., S.M., T.G. performed the experimental investigation. S.M., O.W., T.G. and M.J.P. wrote software to perform the experiment. O.W., T.G., S.M. and M.J.P. performed formal analysis of the results. All authors contributed to the writing and revision of the manuscript. 

\subsection*{Disclosures}The authors declare that they have no competing or non-competing interests.

\subsection*{Data and materials availability:} Additional data and materials will be made available online.

\end{document}